\documentclass{iaus}
\usepackage{graphicx}

\newcommand{\apj}{{\it ApJ}}
\newcommand{\aj}{{\it AJ}}
\newcommand{\mnras}{{\it MNRAS}}
\newcommand{\aanda}{{\it A\&A}}
\newcommand{\pasp}{{\it PASP}}

\title[Young massive star clusters] 
{Young massive star clusters:\\ Achievements and challenges}

\author[Richard de Grijs]   
{Richard de Grijs}
\affiliation{Kavli Institute for Astronomy and Astrophysics, Peking
University, Beijing 100871, China\\ email: {\tt grijs@kiaa.pku.edu.cn}\\[\affilskip]
Department of Physics \& Astronomy, The University of Sheffield, Sheffield S3 7RH, UK\\
email: {\tt R.deGrijs@sheffield.ac.uk} \\[\affilskip]
National Astronomical Observatories, Chinese Academy of Sciences, Beijing 100012, China}
\pubyear{2009}
\volume{266}  
\pagerange{xxx--yyy}
\jname{Star clusters: Basic galactic building blocks throughout time and space}
\editors{R. de Grijs and J. L\'epine, eds.}
\begin{document}

\maketitle

\begin{abstract}
In spite of significant recent and ongoing research efforts, most of
the early evolution and long-term fate of young massive star clusters
remain clouded in uncertainties. Here, I discuss our understanding of
the initial conditions of star cluster formation and the importance of
initial substructure for the subsequent dynamical-evolution and
mass-segregation timescales. I also assess our current understanding
of the (initial) binary fraction in star clusters and the shape of the
stellar initial mass function at the low-mass end in the
low-metallicity environment of the Large Magellanic Cloud. Finally, I
question the validity of our assumptions leading to dynamical cluster
mass estimates. I conclude that it seems imperative that observers,
modellers and theorists combine efforts and exchange ideas and data
freely for the field to make a major leap forward. 
\end{abstract}

\firstsection 
\section{Introduction}

It is now widely accepted that stars do not form in isolation, at
least for stellar masses above $\sim$0.5 M$_\odot$. In fact, 70--90\%
of stars may form in a clustered mode (cf. Lada \& Lada 2003). Star
formation results from the fragmentation of molecular clouds, which in
turn preferentially leads to star cluster formation. Over time,
clusters dissolve or are destroyed by interactions with molecular
clouds or tidal stripping by the gravitational field of their host
galaxy.

A significant amount of recent research has focused on whether at
least some of the young massive star clusters (YMCs) associated with
the most violent starburst events in the local Universe may evolve
into counterparts of the ubiquitous globular clusters (GCs) observed
in almost all local galaxies. The main motivation for these studies
was essentially twofold. First, for any compact, bound cluster to
survive for a cosmologically significant length of time, its stellar
initial mass function (IMF) must have a sufficient number of low-mass
stars (acting as dynamical `glue') to keep it together for so
long. This places strong constraints on the IMF shape of any surviving
YMCs and GCs. Second, as presumed hallmarks of the most violent
galaxy-wide starburst episodes, YMCs trace the star-formation (and, to
some extent, the assembly) histories of their host galaxies. As such,
they serve as the proverbial light houses in the dark.

Rather than regurgitating the well-known issues affecting our detailed
understanding of YMC-to-GC evolution (for reviews see, e.g., de Grijs
\& Parmentier 2007; de Grijs 2010), in this contribution I will focus
on recent results as regards the shape of the low-mass IMF in cluster
environments (Section 2), the effects of the initial conditions on
cluster dynamics (Section 3) and the binary contributions in star
clusters as a function of age (Section 4). In Section 5, I will
conclude this contribution by highlighting some of the remaining
uncertainties hampering a more detailed understanding of the
underlying physics driving star cluster evolution and its
observational interpretation.

\section{The low-mass initial mass function}

The shape of the stellar IMF in the solar neighbourhood for stellar
masses $> 1$ M$_\odot$ has essentially remained unchallenged since
Salpeter's (1955) seminal study. Neverheless, the origin of its
apparent universality is still hotly debated (e.g., Bonnell et
al. 2007; Goodwin \& Kouwenhoven 2009). Better constraining the
physical origin of the IMF will have a major impact on, e.g., our
understanding of the conditions prevailing in a wide range of
starburst events, and the formation of the first stars and clusters in
the early Universe. However, for the latter we would need to follow
the full radiative cooling processes from primordial gas and the
subsequently formed metallic elements in full detail!

At low masses, most current models agree that the solar-neighbourhood
IMF flattens. This can be modelled by either multiple power-law or
lognormal mass distributions (cf. Kroupa 2001; Chabrier 2003). While
the former provides a mathematically convenient and observationally
useful scaling law, the latter is supported by realistic numerical
simulations in an attempt to understand the underlying physics
(Hennebelle \& Chabrier 2008). These simulations take into account
dynamical depletion of the lowest-mass stars and replace the idea of a
single Jeans mass for all newly formed stars in a given molecular
cloud by a distribution of local Jeans masses which are representative
of the lognormal density distribution of the turbulent, fragmenting
gas. As statistically significant samples of roughly coeval stars,
rich young star clusters play a major role in constraining the
low-mass IMF. Open questions remaining in this field relate to, among
others, the initial structure of newly formed clusters and whether the
ubiquitous mass segregation observed in clusters of any age is
dynamical or perhaps primordial (i.e., related to the process of star
formation).

Note, however, that Goodwin \& Kouwenhoven (2009) recently argued that
a universal IMF does not necessarily provide unambiguous information
about the star-formation activity from which the individual stars
originated. One needs to take into account the initial binary fraction
and mass-ratio distribution as well as the core-mass function and
star-formation efficiency. They showed convincingly that very
different (binary) mass-ratio distributions can produce very similar
IMFs from very similar {\it core}-mass functions, while the resulting
IMFs are also (to first order) insensitive to the binary fraction.

Ignoring the Goodwin \& Kouwenhoven (2009) suggestions for the moment,
preliminary clues as to the shape of the low-mass IMF (down to
$\sim$0.15--0.30 M$_\odot$) in the low-metallicity ($Z \sim 0.4$
Z$_\odot$) environment of young ($\sim$4--45 Myr) Large Magellanic
Cloud (LMC) clusters have recently been uncovered on the basis deep
{\sl Hubble Space Telescope} imaging observations (e.g., Da Rio et
al. 2009; Liu et al. 2009a,b). Da Rio et al. (2009) reach a lower-mass
limit of $\simeq 0.43$ M$_\odot$ in constructing their IMF of the LH
95 stellar association, while Liu et al. (2009a,b) push their method
as low as 0.15--0.30 M$_\odot$ in the populous, young ($\sim$20--45
Myr-old) clusters NGC 1805 and NGC 1818. These analyses are
particularly challenging in view of the stellar population's
evolutionary state, requiring pre-main-sequence modelling to derive
robust stellar mass estimates, which is notoriously difficult.

\begin{figure}
\begin{center}
 \includegraphics[width=0.5\columnwidth]{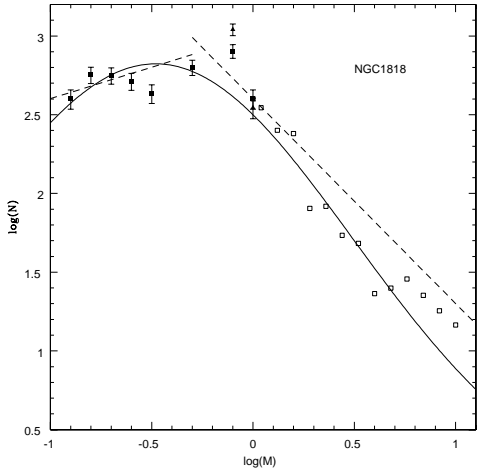}
 \caption{Complete mass function of NGC 1818. The dashed lines show
the standard Kroupa (2001) IMF while the solid line represents the
best-fitting lognormal distribution. See Liu et al. (2009a) for more
details.}
   \label{fig1}
\end{center}
\end{figure}

While these results must therefore be taken with a degree of caution,
they appear to imply that the IMFs of these young clusters are
essentially the same as that in the solar neighbourhood, although the
characteristic stellar masses are somewhat higher. Figure~\ref{fig1}
shows the complete mass function of NGC 1818 (Liu et al. 2009a) down
to the 50\% completeness limit. One would ideally want to probe
younger star-forming regions to reach firmer conclusions, but these
are inevitably obscured by large amounts of dust, hence requiring deep
and often wide-field infrared (IR), (sub)millimetre, radio and X-ray
surveys (as well as pointed observations) that are now coming online
(e.g., the Spitzer Space Telescope's GLIMPSE survey or the UKIRT IR
deep-sky survey, UKIDSS; e.g., Benjamin et al. 2003; Lucas et
al. 2008) and which probe the low-mass stellar mass distribution in
particular (see, e.g., Rathborne et al. 2009).

\section{Early dynamical evolution}

Simulations of star cluster evolution almost always assume that the
stars are initially smoothly distributed and in dynamical
equilibrium. However, both observations and the theory of star
formation tell us that this is not how clusters form. We recently
investigated the effects of substructure and initial clumpiness on the
early evolution of clusters (Allison et al. 2009; see also references
therein). Comparisons with observations will allow us to constrain how
much initial substructure can be present. The most massive stars in
young star clusters are almost always observed to be mass segregated
(e.g., Hillenbrand \& Hartmann 1998; de Grijs et al. 2002a,b,c;
Gouliermis et al. 2004). A crucial question triggered by this
observation relates to the physical origin of this characteristic mass
distribution.  Do massive stars form in the centres of clusters, or do
they migrate there over time due to gravitational interactions with
other cluster members? In smooth, relaxed clusters it has been argued
that the most massive stars must form in the cores (e.g., Bonnell \&
Davies 1998; and references therein), which is therefore often
referred to as primordial mass segregation (but see Ascenso et
al. 2009). But does substructure perhaps play an important role?

Both observational evidence (e.g., Larson 1995; Testi et al. 2000;
Elmegreen 2000; Lada \& Lada 2003; Gutermuth et al. 2005; Allen et al.
2007) and theorical considerations suggest that young star clusters
tend to form with a significant amount of substructure. Their
progenitor molecular clouds are observed to have significant levels of
substructure in both density and kinematics (e.g., Carpenter \& Hodapp
2008), which is likely induced by the supersonic turbulence thought to
dominate molecular cloud structure (e.g., Mac Low \& Klessen 2004;
Ballesteros-Paredes et al. 2007). Observations also support a scenario
in which young clusters lose their substructure on timescales of
$<2$~Myr (e.g., Cartwright \& Whitworth 2004; Schmeja et
al. 2008). Simulations suggest that the only way in which this could
happen is if clusters are born dynamically cool (Goodwin et al. 2004;
Allison et al. 2009). On the basis of these arguments, Allison et
al. (2009) recently performed an ensemble of {\it N}-body simulations
aimed at exploring the earliest phases of cluster evolution. They find
that cool, substructured clusters appear to mass segregate dynamically
for stellar masses down to a few M$_\odot$ on timescales of a few Myr.
This is\break reminiscent of the observational status of the Orion
Nebula Cluster (e.g., Bonnell \& Davies 1998; Allison et al. 2009;
Moeckel \& Bonnell 2009).

Allison et al. (2009) modelled an initially highly substructured
cluster (using multiple realisations to assess the numerical
uncertainties), characterised by a ratio of the kinetic to potential
energy of 0.3 (where 0.5 is virialised) and a fractal dimension of 1.6
(where 3 corresponds to a spherically symmetric distribution),
containing 1000 stars drawn from a Kroupa (2001) IMF spanning the mass
range from 0.08 to 50 M$_\odot$. Given the cluster's nonsphericity, we
used a novel approach to determine the degree of mass segregation,
quantified by using the concept of the `minimum spanning tree'
(MST). A sample's MST corresponds to the path connecting all points in
the sample with the shortest possible path length, and which contains
no closed loops (see, e.g., Prim 1957). We defined a `mass-segregation
ratio',
\begin{equation}
\Lambda_{\rm MSR} = \frac{\langle l_{\rm norm} \rangle}{l_{\rm
massive}} \pm \frac{\sigma_{\rm norm}}{l_{\rm massive}},
\end{equation}
where $\langle l_{\rm norm} \rangle$ is the average length of the MST
of sets of $N_{\rm MST}$ random stars and $l_{\rm massive}$ is the
length of the MST of the $N_{\rm MST}$ most massive stars. The
dispersion associated with the average length of the random MSTs is
roughly Gaussian and can therefore be quantified by the standard
deviation, $\sigma_{\rm norm}$. Using this definition, if
$\Lambda_{\rm MST} > 1$, a given sample of cluster stars is mass
segregated.

\begin{figure}
\begin{center}
 \includegraphics[width=0.5\columnwidth]{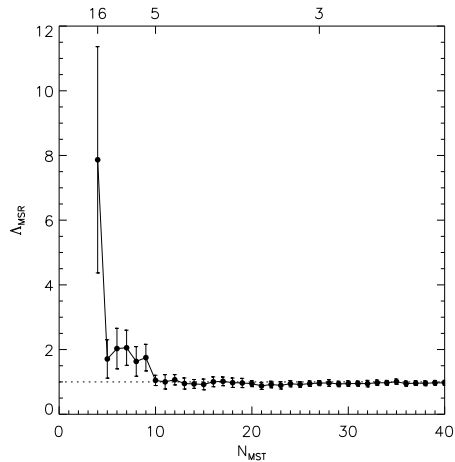}
 \caption{Mass-segregation ratio, $\Lambda_{\rm MST}$, for the Orion
 Nebula Cluster. The dashed line indicates the absence of mass
 segregation. See Allison et al. (2009) for more details.}
   \label{fig2}
\end{center}
\end{figure}

Figure \ref{fig2} shows an application of this method to the 900 stars
in the Orion Nebula Cluster for which masses are available
(Hillenbrand 1997). The method clearly identifies the mass segregation
of the central Trapezium system at $N_{\rm MST} = 4$, $\Lambda_{\rm
MST} = 8.0 \pm 3.5$, but it also shows that there appears to be a
secondary level of mass segregation involving the nine most massive
stars $> 5$ M$_\odot$ (cf. Hillenbrand \& Hartmann 1998; see Allison
et al. 2009 for additional details).

More work is required to systematically address the most likely
initial conditions for cluster formation leading to the observed
configurations. In particular, we have thus far not included binary
stars, although they are expected to have an effect on the MST
lengths. If a binary is resolved, it is very likely that the two
components will be linked as a node, as will higher-order
systems. Allison et al. (2009) suggest that this raises the
possibility that MSTs could be very useful in locating binary and
multiple systems by looking for short links within the MST (but see
Cartwright \& Whitworth 2005). In addition, many cluster observations
suffer from significant incompleteness, particularly near the most
massive stars where low-mass stars cannot be detected even if they are
present. This poses a significant problem as it is impossible to know
if there are many low-mass stars in the `central' regions (in whatever
way defined in the presence of substructure).

\section{The contribution of binaries to star cluster evolution}

More often than not, simulations of star clusters neglect the presence
of binary stars. Observations of local star-forming regions lead us to
suspect that all, or nearly all, stars form in binary or triple
systems (Goodwin \& Kroupa 2005; Duch\^ene et al. 2007; Goodwin et al.
2007). Such systems significantly affect the dynamical evolution of
the cluster, yet the initial binary fractions in dense star clusters
are poorly known. Almost all studies of binarity have been limited to
nearby solar-metallicity populations (see Duch\^ene 1999 and Duch\^ene
et al. 2007 for reviews). However, it might be expected that
metallicity (e.g., through its effects on cooling and hence on the
opacity limit for fragmentation) will play a role in the fragmentation
of cores to produce binary systems (Bate 2005; Goodwin et al. 2007).

The binary fractions in more distant, massive clusters have not yet
been studied thoroughly, because of observational limitations. Note,
however, that statistical colour-magnitude analysis based on
artificial-star tests offers a promising alternative (e.g., Romani \&
Weinberg 1991; Rubenstein \& Bailyn 1997; Bellazzini et al. 2002; Cool
\& Bolton 2002; Zhao \& Bailyn 2005; Davis et al. 2008). In addition,
all clusters thus far studied in this way are old stellar systems, in
which dynamical evolution is expected to have altered the initial
binary population significantly. Efforts have begun to address this
issue for the much more distant young populous clusters in the LMC
(e.g., Elson et al. 1998). Hu et al. (2009) estimate that the binary
fraction in NGC 1818 in the mass range between 1.3 and 1.6 M$_\odot$
is $\sim$0.35 for systems with an approximately flat mass-ratio
distribution, $q$, for $q>0.4$. This is consistent with a {\it total}
binary fraction of F stars of 0.6 to unity. Note, however, that in
view of recent developments and the discovery of multiple main
sequences in a variety of star clusters (see, e.g., Piotto, this
volume), this may need to be revised downwards.

\begin{figure}
\begin{center}
 \includegraphics[width=0.5\columnwidth]{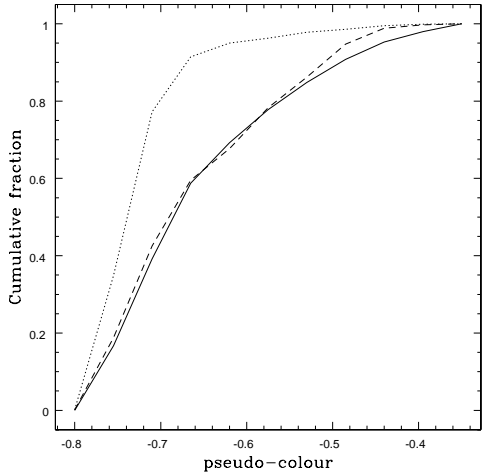}
 \caption{Observed cumulative distribution function with pseudo-colour
in NGC 1818 (solid line; full stellar sample) compared with an
artificial stellar population with zero binary fraction ($f_{\rm b}$,
dotted line), and the best-fitting (r.m.s.) uniform mass-ratio
distribution of $f_{\rm b} = 0.62 \pm 0.05$ (dashed line). See Hu et
al. (2009) for more details.}
   \label{fig3}
\end{center}
\end{figure}

In Fig. \ref{fig3} (Hu et al. 2009), I show the cumulative
distribution function (CDF) with pseudo-colour (defined as the colour
change along the main sequence in the colour-magnitude diagram) as
observed for the entire magnitude range of interest in NGC 1818 (solid
line) and for a stellar population with photometric errors but no
binaries (dotted line). This is clearly a very bad fit to the
data. The figure also shows the best-fitting (r.m.s.) binary fraction
($f_{\rm b}$), assuming a uniform mass-ratio distribution, of $f_{\rm
b} = 0.62 \pm 0.05 (1\sigma)$ (dashed line). Note that the fit is poor
at larger pseudo-colours. This is always the case and is due to the
presence of unresolved higher-order systems.

The combination of the results presented in this and the previous
sections triggers a number of questions. Do high binary fractions
affect mass segregation at early times or the relaxation of
substructure? Do they leave observational signatures? NGC 1818 is
several crossing times old, so that the binary population should have
been modified by dynamical interactions. In particular, soft (i.e.,
wide) binaries are expected to have been destroyed by this
age. Therefore, the high binary fraction found for F stars suggests
that these binaries are relatively `hard' and able to survive
dynamical encounters.

\section{Uncertainties abound}

For any cluster, but particularly for the lower-mass end of the
cluster mass function, it seems clear that the effect of binaries,
mass segregation and the dynamical alteration of mass functions by
two-body relaxation are important factors that cannot be ignored.

We recently explored the usefulness of the diagnostic age versus
mass-to-light-ratio diagram in the context of YMCs and Galactic open
clusters (e.g., Smith \& Gallagher 2001; Mengel et al. 2002; McCrady
et al. 2003, 2005; Larsen et al. 2004; Bastian et al. 2006; Goodwin \&
Bastian 2006; Moll et al. 2008; see de Grijs \& Parmentier 2007 for a
review). This diagram is often used in the field of extragalactic
young to intermediate-age massive star clusters to constrain the shape
of their stellar IMF, as well as their stability and the likelihood of
their longevity. Based on high-resolution spectroscopy to obtain the
objects' integrated velocity dispersions, $\sigma$, and on
high-spatial-resolution imaging to obtain accurate projected
half-light radii, $r_{\rm hl}$, most authors then apply Spitzer's
(1987) equation,
\begin{equation}
\label{spitzer.eq}
M_{\rm dyn} = \eta \frac{r_{\rm hl} \sigma^2}{G} ,
\end{equation}
to obtain the dynamical cluster masses, $M_{\rm dyn}$ ($G$ is the
gravitational constant and $\eta \approx 9.75$ is a dimensionless
parameter which is usually assumed to be constant; but see Fleck et
al. 2006; Kouwenhoven \& de Grijs 2008).

Despite a number of simplifying assumptions (see, e.g., de Grijs \&
Parmentier 2007 for a review; Moll et al. 2008), one can get at least
an initial assessment as to whether a given cluster may be (i)
significantly out of virial equilibrium, in particular `super-virial',
(ii) significantly over- or underabundant in low-mass stars, or (iii)
populated by a significant fraction of binary and higher-order
multiple systems. This has led to suggestions that, in the absence of
significant external perturbations, young massive clusters (YMCs)
located in the vicinity of the simple stellar population models and
aged $\ge 10^8$ yr may survive for a Hubble time and eventually become
old GC-like objects (e.g., Larsen et al. 2004; Bastian et al. 2006; de
Grijs \& Parmentier 2007).

Using the massive young Galactic cluster Westerlund~1 as a key
example, we cautioned that stochasticity in the IMF introduces
significant additional uncertainties (de Grijs et al. 2008).  For
Galactic open clusters, the effect of binaries within clusters may
well account for most of the displacement of the observed cluster
positions to below the model curves in de Grijs et
al. (2008). Kouwenhoven \& de Grijs (2008) pointed out that if the
velocity dispersion of binary systems were similar to the velocity
dispersion of the cluster as a whole, the {\it observationally
measured} velocity dispersion would overestimate the mass of a
cluster. Based on a comparison with Kouwenhoven \& de Grijs (2008), in
de Grijs et al. (2008) we concluded that the vast majority of our
sample of open clusters are indeed expected to be binary dominated.

We also note that the cluster masses may well have been overestimated
by factors of a few through the universal use of
Eq. (\ref{spitzer.eq}). In particular, for highly mass-segregated
clusters containing significant binary fractions, a range of stellar
IMF representations, and for combinations of characteristic relaxation
timescales and cluster half-mass radii, the adoption of a single
scaling factor $\eta \approx 9.75$ introduces systematic offsets. To
correct for these, we would need to adopt smaller values of $\eta$
(e.g., Fleck et al. 2006; Kouwenhoven \& de Grijs 2008), and this
would thus lead to dynamical mass overestimates if $\eta = 9.75$ were
assumed.

We have now reached a stage in star cluster studies in which it is
imperative to combine observational, theoretical and numerical
modelling efforts to make the next major leap in our physical
understanding. The level of detail required to make significant
progress necessitates the combined forces of modellers in the fields
of stellar population synthesis, {\it N}-body simulations and
smooth-particle hydrodynamics, as well as observers with a good grasp
of the intricacies of curent-generation data problems and reduction
issues. It looks, therefore, that the conventional small research
teams of the past may soon need to expand and become more inclusive to
make significant and exciting headway.

\begin{acknowledgements}
I thank the Scientific Organising Committee of IAU Symposium 266 for
their vote of support to deliver this review talk. I would also like
to highlight that most of the results presented in this contribution
are based on PhD thesis work by Richard Allison (University of
Sheffield), Qiang Liu and Yi Hu (National Astronomical Observatories,
Chinese Academy of Sciences). I acknowledge funding to attend this
meeting from the Royal Society, the International Astronomical Union
and the Learned Societies Fund of the University of Sheffield.
\end{acknowledgements}

\end{document}